\documentclass[%
 reprint,
 amsmath,amssymb,
 aps,
]{revtex4-2}

\usepackage{multirow}
\usepackage{diagbox}
\usepackage{graphicx}% Include figure files
\usepackage{dcolumn}% Align table columns on decimal point
\usepackage{bm}% bold math
\begin{document}

\preprint{APS/123-QED}

\title{Enantiosensitive exceptional points in open chiral systems}% Force line breaks with \\

\author{Nicola Mayer$^{1,2}$, Alexander L{\"o}hr$^{1}$, Nimrod Moiseyev$^{3,4}$, Misha Ivanov$^{1,3,5,6}$ and Olga Smirnova$^{1,3,7}$}
\affiliation{%
 $^1$Max-Born-Institut, Max-Born-Str. 2A, Berlin, 12489 Germany\\
 $^2$ Attosecond Quantum Physics Laboratory, Physics Department, King's College London, Strand, London WC2R 2LS, UK\\
 $^3$ Technion-Israel Institute of Technology, 3200003, Haifa, Israel\\
 $^4$ Institute of Advanced Studies in Theoretical Chemistry, Technion-Israel Institute of Technology, 3200003, Haifa, Israel\\
 $^5$ Department of Physics, Humboldt Universit\"at zu Berlin, Newtonstr. 15, Berlin, D-12489, Germany\\
 $^6$ Department of Physics, Imperial College London, London, United Kingdom\\
 $^7$ Technische Universit\"at Berlin, Straße des 17.\ Juni 135, Berlin, 10623, Germany
}%

\date{\today}% It is always \today, today,

\begin{abstract}
Exceptional points (EPs) are remarkable spectral degeneracies in a non-Hermitian system's parameter space, where both eigenvalues and eigenstates coalesce. Here, we show that in non-Hermitian molecular chiral systems the position of EPs in the parameter space is enantiomer-specific. First, we show that encircling the EP of one enantiomer drives robust topological population transfer in the chiral molecule while its mirror twin remains unaffected, offering a new route for selective chiral control. Second, we reveal how resonant excitation of EPs in chiral molecules can amplify weak chiral effects, offering an alternative approach to the enhancement of chiral interactions. Third, we demonstrate that a twisted chiral fiber immersed in a liquid solution of chiral molecules exhibits topologically different behavior depending on the solution's enantiomeric excess, offering a new approach to the detection of molecular chirality. Our results combine high enantiosensitivity with topological robustness in chiral discrimination and control, paving the way for new approaches in the exploration of non-Hermitian and chiral phenomena.
\end{abstract}

\maketitle

%\tableofcontents

\section{Introduction}
The study of non-Hermitian systems, characterized by Hamiltonians that do not commute with their self-adjoints $H\neq H^{\dagger}$, has revealed new topological phenomena in non-conservative dynamics shaped by gain, loss and non-reciprocity \cite{moiseyev_2011,Moiseyev2018,Feng:2017aa,Ozdemir:2019aa}. Non-Hermitian systems are central in the study of parity-time ($\mathcal{PT}$) symmetry in both quantum and classical mechanics \cite{Bender:2019aa}. An Hamiltonian is said to be $\mathcal{PT}$-symmetric if it 
commutes with the combined action of the parity $\mathcal{P}$, which performs spatial inversion, and the time-reversal $\mathcal{T}$ operator, which performs complex conjugation and time inversion. Remarkably, a large class of non-conservative systems described by a non-Hermitian Hamiltonian with balanced gain and loss can respect such symmetry, leading to a purely real eigenspectrum that is said to be in the $\mathcal{PT}$-symmetric topological phase until a symmetry-breaking threshold is reached. In the $\mathcal{PT}$-broken phase, the non-Hermitian Hamiltonian exhibits instead complex eigenvalues, leading to differential attenuation or gain of the corresponding eigenstates. The $\mathcal{PT}$-symmetric and $\mathcal{PT}$-broken topological phases of a non-Hermitian system are connected in the parameter space by the so-called exceptional points (EPs) \cite{Ozdemir:2019aa,Feng:2017aa}, spectral singularities where the complex eigenvalues and corresponding eigenstates of the non-Hermitian Hamiltonian coalesce \cite{Heiss:1999,Gao:2015aa,PhysRevLett.86.787}, leading to non-trivial behavior in their vicinity that has no counterpart in Hermitian systems.

The combination of chiral, $\mathcal{P}$-breaking, media with $\mathcal{PT}$-systems tuned to an exceptional point has recently sparked a series of exciting research efforts \cite{DeCorte:2024aa,Katsantonis:2022aa,Katsantonis:2020ab,Katsantonis:2020aa,Droulias:2019aa,DeCorte:2024ab,PhysRevA.105.053711,PhysRevLett.124.083901,Wu:2022aa}. These studies span a wide range of platforms and phenomena, from chiral metamaterials enabling advanced polarization control \cite{Droulias:2019aa,Katsantonis:2020aa,Katsantonis:2020ab,DeCorte:2024aa}, to chiral waveguides displaying $\mathcal{PT}$ effects in absence of gain/loss \cite{DeCorte:2024ab}. The interplay of non-Hermitian and chiral effects offers a powerful novel framework with implications for light manipulation, topological photonics and sensing.

In parallel, recent years have seen major advances in chiral light-matter interactions involving chiral molecules \cite{science_new_age}. Whereas standard methods like absorption circular dichroism or optical rotation are limited by the weak interaction of the magnetic component of light with molecules, new ‘electric-dipole revolution’ \cite{Ayuso:2022aa} techniques bypass the need for magnetic dipole interactions and can enhance the chiral signals by several orders of magnitude \cite{Ayuso:2019aa,Khokhlova:2022aa,Rego:2023aa,Ordonez:2023ab,Vogwell:2023aa,Ayuso:2021aa,Habibovic:2024aa}. The increase in efficiency of light-matter coupling is dramatic for optical and lower (e.g.\,THz or microwave) frequencies and medium-sized molecules, such as chiral molecules in living matter \cite{Zheng:2023aa}. Developing from this research field, novel topological approaches that rely on the geometric nature of chirality have opened access to topological enantiosensitive observables that endow robustness to these methods against experimental noise \cite{Ordonez:2019ab,Ordonez:2023aa,Mayer:2024aa,Schwennicke:2022aa,Peter:2024aa}.

In this work, we combine the developments in chiro-optical spectroscopy and the role of $\mathcal{P}$-breaking in non-Hermitian systems to explore a new approach to chiral molecular sensing that leverages the sensitivity of the position of EPs to the handedness of molecular enantiomers. We identify this type of EPs as ``enantiosensitive EPs".

First, we merge the electric-dipole techniques \cite{Ayuso:2022aa} with the non-Hermitian formalism, achieving topological enantiosensitive population transfer that relies on purely electric-dipole transitions to encircle an enantiosensitive EP in a chiral molecule coupled to the photoionization continuum. Then, we show that even when weak chiral interactions employing magnetic dipoles occur, the presence of enantiosensitive EPs can lead to an enhancement of chiral dichroism in the decay of chiral metastable resonances. Finally, exploiting the nature of EPs as transition points between the $\mathcal{PT}$-broken and $\mathcal{PT}$-symmetric regimes of a system, we propose a new approach to detect the concentration of right- or left-handed molecules in a liquid solution using a non-Hermitian chiral twisted fiber tuned to an enantiosensitive EP, showing that small fluctuations of enantiomeric excess can be detected with high sensitivity by involving the topological transition between the two $\mathcal{PT}$ regimes. 

\begin{figure*}[ht!]
\begin{center}
\includegraphics[width=17cm, keepaspectratio=true]{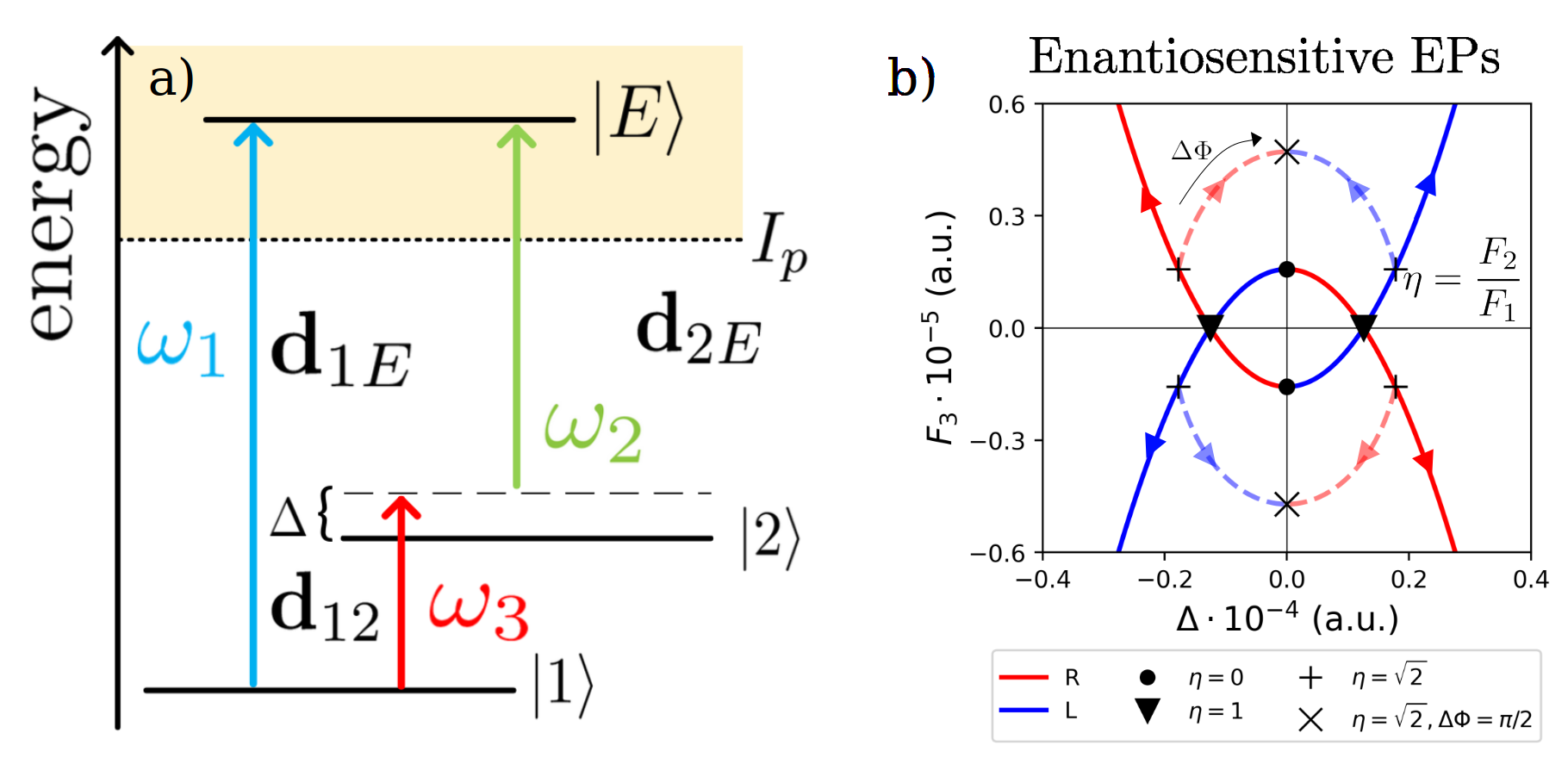}
\caption{\textbf{Enantiosensitive exceptional points of a non-Hermitian chiral molecule.} \textbf{(a)} Three-level model representing two bound states $|1\rangle$ and $|2\rangle$  of a chiral molecule coupled to each other and the continuum $|E\rangle$ via electric-dipole transitions by a three-color laser field with frequencies $\omega_1=\omega_2+\omega_3$. \textbf{(b)} Position of the EPs in the parameter space $(\Delta,F_3)$ for varying ratios $\eta=F_2/F_1$ and fixed $F_1$. The solid red (blue) lines correspond to the right (left) enantiomer, where the arrows indicate the directions along which the EPs move for increasing $\eta$. The black circles (triangles) correspond to the position of the EPs for $\eta=0$ ($\eta=1$), where enantiosensitivity is lost. The red (blue) circles correspond to the position of the EPs for $\eta=\sqrt{2}$ for the right (left) enantiomer. The dashed red (blue) lines show how the EPs of the right (left) enantiomer move for fixed $\eta=\sqrt{2}$ and varying relative phase $\Delta\Phi\in[0,\pi/2]$ of the three-photon matrix element $\Omega_{123}$. At $\Delta\Phi=\pi/2$ the enantiosensitivity is lost and the EPs of the two enantiomers are on the vertical axis $\Delta=0$.}
\label{Fig1}
\end{center}
\end{figure*}
\begin{figure*}
\begin{center}
\includegraphics[width=18cm, keepaspectratio=true]{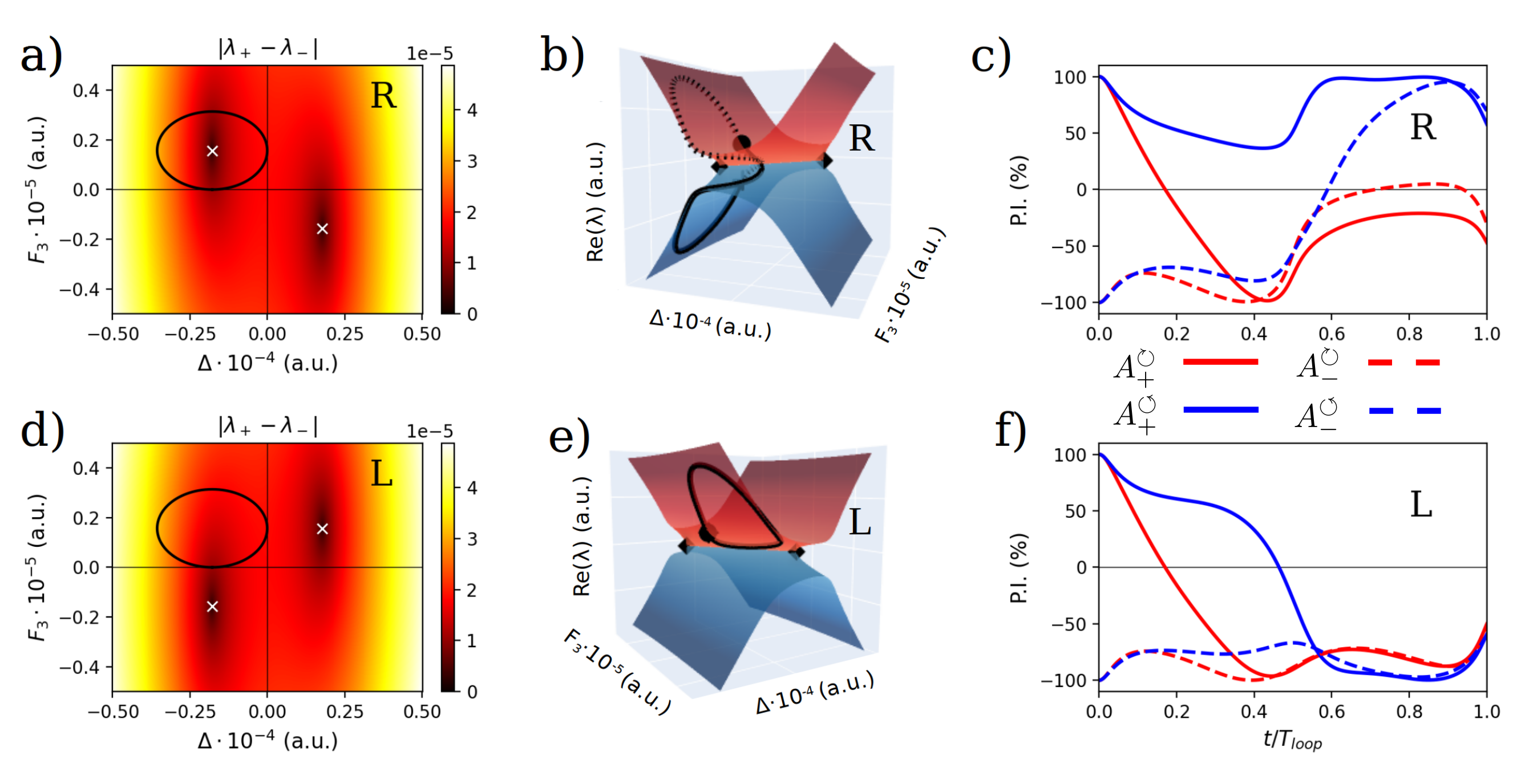}
\caption{\textbf{Enantiosensitive topological population transfer in non-Hermitian chiral molecules.} \textbf{(a,d)} Absolute value of the difference in quasi-energies $|\lambda_+-\lambda_-|$ for the right (\textbf{a}) and left (\textbf{d}) enantiomer in the $(\Delta,F_3)$ parameter space. The black solid line shows the path enclosing one of the exceptional points (white cross) of the right enantiomer. \textbf{(b,e)} Trajectories of the adiabatic solutions for the path in (a,d) for the right (\textbf{b}) and left (\textbf{e}) enantiomers on the quasi-energy surfaces $\mathrm{Re}[\lambda_\pm(\Delta,F_3)]$. Solid (dashed) lines correspond to clockwise (counter-clockwise) encirclement. The right enantiomer in (b) shows the adiabatic flip associated to the encirclement of an EP, leading to a transfer of population from one adiabatic state to the other at the end of the loop, while for the left enantiomer in (e) the initial and final population at the end of the loop coincide. \textbf{(c,f)} Time-dependent populations population inversion $A^{\circlearrowright/\circlearrowleft}_{\pm}(t)$ in the adiabatic states for a dynamical evolution along the path shown in (a,d) for the right (\textbf{c}) and left (\textbf{f}) enantiomer. Red (blue) color corresponds to clockwise (counter-clockwise) encirclement, while solid (dahsed) lines correspond to an initial population in the $|\phi_+\rangle$ ($|\phi_-\rangle$) adiabatic state. The asymmetric enantiosensitive switch is observed for the right enantiomer in (c), where the sign of the population inversion parameter is determined by the sense of encirclement, while for the left enantiomer in (d) the population at the end of the evolution is predominantly in the $|\phi_-\rangle$ regardless of the sense of encirclement and initial conditions.}
\label{Fig2}
\end{center}
\end{figure*}
\section{Enantiosensitive topological population transfer}
A three-color field where
the three polarization vectors $\mathbf{F}_i(\omega_i)$ ($i=1,2,3$) are non-collinear and $\omega_1=\omega_2+\omega_3$ \cite{PhysRevLett.87.183002,Eibenberger:2017aa,Patterson:2013aa}, 
is locally (temporally) chiral \cite{science_new_age,Ayuso:2019aa}. Its
chirality is encoded in the Lissajous figure drawn by the tip
of the total electric field vector as the field
evolves in time. 
Locally chiral light can be used to control enantiosensitive population transfer in randomly oriented ensembles of chiral molecules, as recently demonstrated experimentally 
in the microwave region for rotational states \cite{Eibenberger:2017aa, Perez:2017aa, Eibenberger:2022aa,Leibscher:2019aa}.
It can also induce topological frequency conversion by taking advantage of the diabolic points in the parameter space defined by the modulated, locally chiral, microwave pulses \cite{Schwennicke:2022aa}. All these methods rely on the enantiosensitivity of cyclic population transfer \cite{PhysRevLett.87.183002} between three states of a chiral molecules. Crucially, they are Hermitian in nature, assuming no loss of population. What happens when a chiral molecule interacting with such light is open to the environment, leading to non-conservative dynamics described by a non-Hermitian Hamiltonian?

Starting from the three-level model of a chiral molecule coupled via electric dipole transitions, we introduce losses by placing the upper state of the three-level system in a continuum as shown in Fig. \ref{Fig1}a, representing e.g. photoionization or photodissociation.

The open three-level system can be recast into an equivalent form (see Supplementary Information (SI) \cite{SuppNotes}\nocite{Fedorov:book,PhysRevA.92.052124,Lebedev1975,PhysRevX.8.021066} for the derivation) corresponding to a dissipative two-level system, described by the non-Hermitian Hamiltonian
\begin{equation}
\label{eq:NH0}H=\begin{bmatrix}-\frac{\Delta}{2}-\text{i}\frac{\gamma}{2} & V_{12} \\ V_{21} & \frac{\Delta}{2}+\text{i}\frac{\gamma}{2}
\end{bmatrix}.
\end{equation}
Here $\gamma=(\Gamma_1-\Gamma_2)/2$, where $\Gamma_i=2\pi|\mathbf{d}_{iE}\cdot\mathbf{F}_i|^2$ are the decay rates of the bound state $|i\rangle$ into the continuum $|E\rangle$ with energy $E=E_1+\omega_1$ where $\mathbf{d}_{iE}=\langle i|\mathbf{d}|E\rangle$ are the corresponding transition matrix elements, and $\Delta=E_2-E_1-\omega_3$ is the detuning of the one-photon dipole transition $\mathbf{d}_{12}$ between the two bound states driven by the field $\mathbf{F}_3$ (see Fig. \ref{Fig1}a).
The off-diagonal couplings are
\begin{eqnarray}V_{12}&=&-2\pi\text{i}\left(\mathbf{d}_{1E}\cdot\mathbf{F}_1\right)^*\left(\mathbf{d}_{2E}\cdot\mathbf{F}_2\right)+\mathbf{d}_{12}\cdot\mathbf{F}_3,\\
V_{21}&=&2\pi\text{i}\left(\mathbf{d}_{1E}\cdot\mathbf{F}_1\right)\left(\mathbf{d}_{2E}\cdot\mathbf{F}_2\right)^*+\left(\mathbf{d}_{21}\cdot\mathbf{F}_3\right)^*.\end{eqnarray}
The first term on the right-hand side corresponds to the two-photon Raman-like coupling through the continuum (see Fig. \ref{Fig1}a), while $\mathbf{d}_{12}\cdot\mathbf{F}_3$ is the Rabi frequency of the one-photon coupling. %A change in molecular handedness inverts all transition dipoles $\mathbf{d}^{R}=-\mathbf{d}^L$, leaving all terms of the Hamiltonian unchanged except for the one-photon Rabi frequency, leading to the enantiosensitive effects explored below.

The eigenvalues of the non-Hermitian Hamiltonian in Eq. [\ref{eq:NH0}] are 
\begin{equation}\label{eq:lambda_encirc}\lambda_{\pm}=\pm\sqrt{\delta},\end{equation}
where the real part of $\delta$ is controlled by the average decay rate $\Gamma=(\Gamma_1+\Gamma_2)/2$ and is equal for both enantiomers
\begin{eqnarray}\label{eq:realdisc}\mathrm{Re}\left(\delta\right)&=&\frac{\Delta^2}{4}-\frac{\Gamma^2}{4}+\left(\mathbf{d}_{12}\cdot\mathbf{F}_3\right)^2.
\end{eqnarray} 
The imaginary part of $\delta$ 
\begin{eqnarray}
\label{eq:disc}
\mathrm{Im}\left(\delta\right)&=&\gamma\Delta+|\Omega_{123}|\cos(\Delta\Phi),
\end{eqnarray}
where
\begin{eqnarray}
\label{eq:O123}
\Omega_{123}&=&2\pi\left(\mathbf{d}_{1E}\cdot\mathbf{F}_1\right)^*\left(\mathbf{d}_{2E}\cdot\mathbf{F}_2\right)\left(\mathbf{d}_{12}\cdot\mathbf{F}_3\right)\nonumber\\
&=&|\Omega_{123}|\exp(\text{i}\Delta\Phi),
\end{eqnarray}
is enantiosensitive if three contributing dipoles are non-coplanar. 
It is closely related to the enantiosensitive three-photon matrix element governing cyclic transitions of closed (Hermitian) chiral systems \cite{PhysRevLett.87.183002,Eibenberger:2017aa,Patterson:2013aa,Schwennicke:2022aa}.  After averaging over molecular orientations, $\Omega_{123}$ presents the product of molecular pseudoscalar $\left[\mathbf{d}^*_{1E}\times\mathbf{d}_{2E}\right]\cdot\mathbf{d}_{12}$ and light pseudoscalar $\left[\mathbf{F}^*_{1}\times\mathbf{F}_{2}\right]\cdot\mathbf{F}_{3}$ known as chiral correlation function \cite{Ayuso:2019aa} characterizing the handedness of locally chiral light. It shows that both the three molecular dipoles  and the three field polarizations (Fig. \ref{Fig1}a.) must be non-coplanar to achieve enantiosensitive coupling between the molecule and light in electric dipole approximation. Assuming that this condition is met, we can write the phase
$\Delta\Phi$ as a sum of molecular-only (indicated by the subscript $\mathrm{M}$) and field-only (subscript $\mathrm{l}$) parts $\Delta\Phi=\Delta\Phi_{\mathrm{l}}+\Delta\Phi_{\mathrm{M}}$, where $\Delta\Phi_{\mathrm{l}}=\arg[F_1]-\arg[F_2]-\arg[F_3]$ are the relative phases of the three frequencies and $\Delta\Phi_{\mathrm{M}}=\arg[d_{1E}]-\arg[d_{2E}]-\arg[d_{12}]$ are the relative phases of the dipole matrix elements of the molecule, projected along the field's polarization axes. These two terms describe respectively the handedness of the three-color locally chiral field and of the molecule. A change in handedness of either corresponds to a phase shift $\Delta\Phi^R_{l,m}=\Delta\Phi^L_{l,m}+\pi$, leading to a change of sign of $\mathrm{Im}(\delta)$ in Eq. [\ref{eq:disc}].

We now find the positions of EPs in the parameter space 
$(\Delta,F_3)$ defined by the detuning $\Delta$ and the strength $F_3$ of the field driving the one-photon transition at frequency $\omega_3$. We do so by requiring that the real and imaginary part of the discriminant $\delta$ in Eqs. \ref{eq:realdisc} and \ref{eq:disc} to be zero. These are enantiosensitive and given in general by
\begin{eqnarray}
\label{eq:EP}
\Delta^{EP_R}&=&\mp\Delta^{EP_L}=\mp\frac{2\kappa\Gamma}{\sqrt{\kappa^2+4\gamma^2|\mathbf{d}_{12}\cdot\mathbf{e}_3|^2}},\nonumber\\
F_3^{EP_R}&=&F_3^{EP_L}=\mp\frac{\gamma\Gamma}{\sqrt{\kappa^2+4\gamma^2|\mathbf{d}_{12}\cdot\mathbf{e}_3|^2}},
\end{eqnarray}
where $\kappa=2|\Omega_{123}F^{-1}_3|\cos(\Delta\Phi)$. Without loss of generality, we can fix the phase of the laser field in such a way that for one of the two enantiomers $\Delta\Phi=0$ and obtain the simpler expressions
\begin{eqnarray}
\label{eq:EP}
\Delta^{EP_R}&=&\mp\Delta^{EP_L}=\mp2\sqrt{\Gamma^2-\gamma^2},\nonumber\\
F_3^{EP_R}&=&F_3^{EP_L}=\mp\frac{\gamma}{|\mathbf{d}_{12}\cdot\mathbf{e}_3|}.
\end{eqnarray}
It is clear that the positions of the EPs in the parameter space $(\Delta, F_3)$ can be tuned, for example by changing the field strengths $F_1$ and $F_2$, and correspondingly the decay rates $\Gamma_1$ and $\Gamma_2$. This is shown in Fig. \ref{Fig1}b, where we report the positions of the EPs of the molecular enantiomers for a varying ratio $\eta=F_2/F_1$ for fixed $F_1$. 

When $\eta=0$ ($\Gamma_2=0$), the EPs of both enantiomers are on the $\Delta=0$ axis and are not enantiosensitive (black dots in Fig. \ref{Fig1}b). For $\eta>0$, the EPs positions become enantiosensitive, drawing two ``parabolic trajectories" moving in opposite directions for opposite molecular handedness (solid red and blue lines for right and left enantiomers respectively). At $\eta=1$ ($\Gamma_1=\Gamma_2)$ the EPs merge again on the $F_3$ axis (black triangles in the figure); for $\eta>1$ the EPs of the two enantiomers again separate in the parameter space. We note that EPs positions can also be tuned by changing the relative phase $\Delta\Phi$ for fixed $\eta>0$ and $\eta\neq1$ (dashed solid lines in Fig. \ref{Fig1}b for $\eta=\sqrt{2}$). Switching the handedness of the chiral field $\Delta\Phi_l$ flips the sign of  $\Omega_{123}$, swapping the trajectories of the EPs of one enantiomer into the other in Fig. \ref{Fig1}b.

Now that we have demonstrated the natural enantiosensitivity of EPs, we can exploit this property to achieve enantiosensitive topological population transfer in a chiral system. To do so, we use the adiabatic flip \cite{Heiss:1999,moiseyev_2011} and asymmetric switch effect \cite{KAPRALOVAZDANSKA2022168939}. Both effects arise when an EP is encircled in the parameter space. In the adiabatic flip effect, adiabatic evolution along the enclosing path leads to a swap of the adiabatic states. In the asymmetric switch effect, dynamical evolution leads to a population transfer where the final state depends on the sense of encirclement. Both effects are observed in non-Hermitian physics, from gas phase atomic and achiral molecular media \cite{Uzdin:2011aa,Gilary:2012,Zdanska:2014aa,PhysRevA.88.010102}, to non-Hermitian photonic and optical platforms \cite{Doppler:2016aa,Xu:2016aa}. 

To make the topological population transfer enantiosensitive, we choose a path in parameter space that encircles only one of the EPs of the right-handed molecule (see Figs. \ref{Fig2}a,b), where the starting time $t_0$ is chosen such that $F_3(t_0)=F_3(t_0+T_{\mathrm{loop}})=0$ (see Appendix A for exact details on the chosen path). When the evolution is adiabatic, the non-trivial regime realized for the right-handed molecule results in the adiabatic flip effect and the population is transferred from one adiabatic state to the other (Fig. \ref{Fig2}b). For the left enantiomer the evolution is trivial and the population returns to its initial adiabatic state (Fig. \ref{Fig2}e).

When the evolution is dynamical, we quantify the dynamics via the time-dependent normalized population inversion parameters $A^{\circlearrowleft/\circlearrowright}_\pm(t)$, where the superscript indicates the sense of encirclement and the subscript the initially populated adiabatic states. The population inversion parameter is defined for a given loop as
\begin{equation}
    A(t)=\frac{P_+(t)-P_-(t)}{P_+(t)+P_-(t)},
\end{equation}
when $A(t)<0$ ($A(t)<0$), the population is found mostly in the $|\phi_-\rangle$ ($|\phi_+\rangle$) state.
The time-dependent population inversions for the right and left enantiomers are shown in Figs. \ref{Fig2}c and \ref{Fig2}f respectively, where we choose a loop time $T_{\mathrm{loop}}=3\cdot10^{5}$ a.u. (7.2 ps). We note that in the dynamical evolution, the degree of enantiosensitivity of the population transfer -- how many right-handed molecules are found in a given adiabatic state compared to the left-handed molecules -- is a trade off between the slow, adiabatic nature of the topological cycle $T_{\mathrm{loop}}\Delta E>1$ (where $\Delta E$ is the transition frequency between the adiabatic states) and the natural desire to minimize the overall losses, which dictates opposite optimal condition $T_{\mathrm{loop}}\Gamma<1$. For the chosen loop time, the final resi\-dual population in the molecular enantiomers is around $1\%-10\%$. To address wether the enantiosensitive population transfer was successful, we can analyze the population left in the enantiomers at the end of the loop reported in table \ref{tab:respop}. For the right enantiomer, we observe the asymmetric switch effect: the final state on which the population is mostly transferred at the end of the loop depends purely on the sense of encirclement and is independent of the initially populated adiabatic state. For the left enantiomer instead the final population is mostly found in the $|\phi_-\rangle$ adiabatic state regardless of the sense of encirclement and the initial population. Following Ref. \cite{Feilhauer:2020aa}, we can further quantify the asymmetric switch effect by combining the population inversion parameters in a single parameter $\alpha\in[-1,1]$ (see Appendix A for the definition), where $\alpha<0$ indicates a successful asymmetric switch effect. We find that for the right enantiomer $\alpha\simeq-0.25$ while for the left enantiomer $\alpha\simeq0$, indicating an enantiosensitive asymmetric switch population transfer in the molecular enantiomers for same dynamical evolution. 

\begin{figure}
\begin{center}
\includegraphics[width=8cm, keepaspectratio=true]{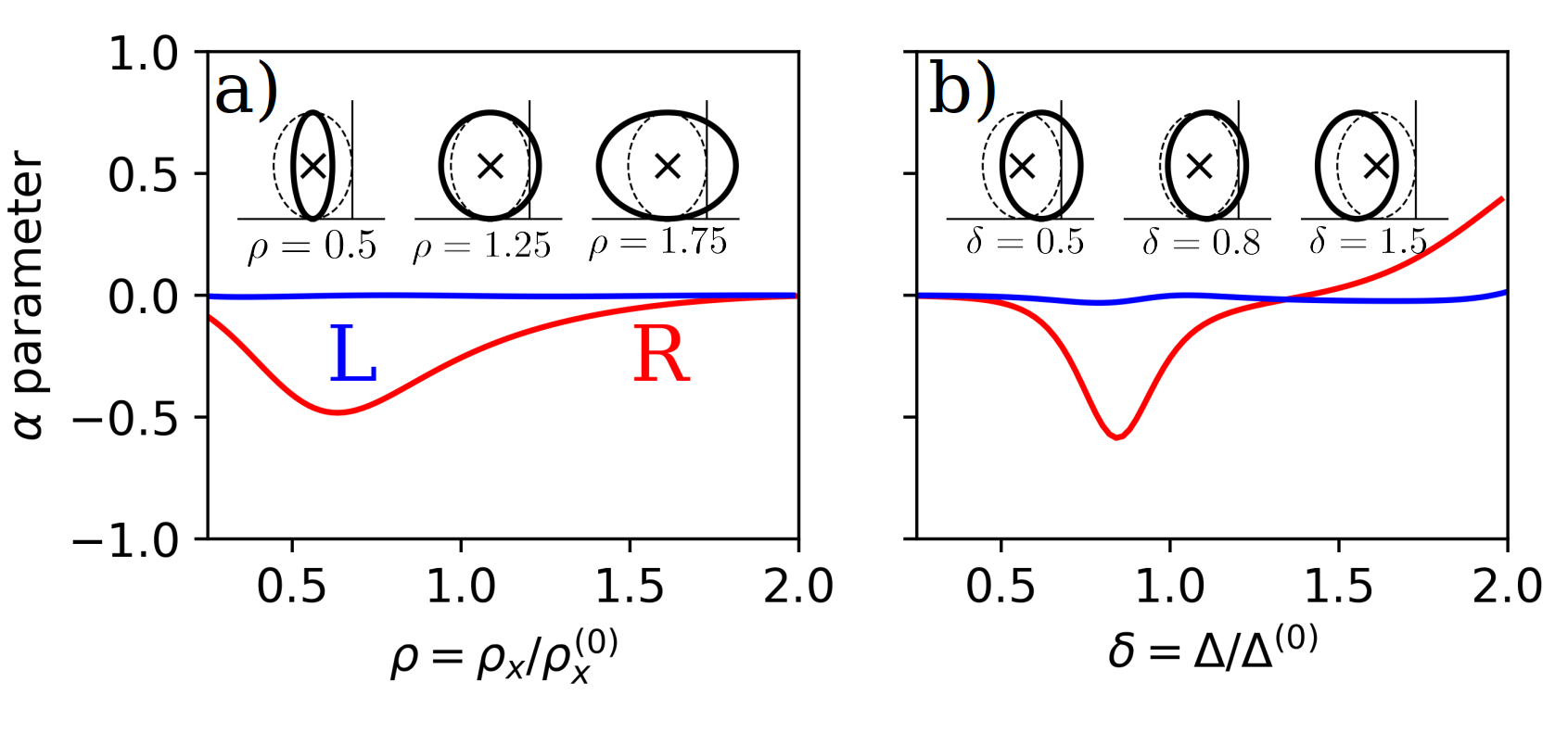}
\caption{\textbf{Stability of the enantiosensitive asymmetric switch effect.} \textbf{(a)} $\alpha$ parameter for paths with a deformed radius $\rho_x=\rho\cdot\rho_x^{(0)}$, where $\rho_x^{(0)}$ is the radius of the path along the $\Delta$ coordinate of the path in Figs. \ref{Fig2}a,d). \textbf{(b)} $\alpha$ parameter for paths with a shifted center $\Delta=\delta\cdot\Delta^{(0)}$, where $\Delta^{(0)}$ is the center of the path in Figs. \ref{Fig2}a,d). The red and blue lines correspond respectively to the right and left enantiomer. The insets show the deformed paths for different values of $\rho$ and $\delta$, where the dotted line shows the undeformed path and the black cross indicates the position of the EP of the right enantiomer.}
\label{Fig3}
\end{center}
\end{figure}

In an experimental realization of our proposal fluctuations in laser parameters, such as intensity or frequency, can lead to a deformation of the path encircling the EP. Yet, because this is a topological phenomenon, the enantiosensitive population transfer is generally stable against such deformations as long as the EP is enclosed by the deformed path. To investigate the stability, we consider paths that have either a deformed radius along the $\Delta$ coordinate $\rho_x=\rho\cdot\rho^{(0)}_x$, or a shifted center $\Delta=\delta\cdot\Delta^{(0)}$, where $\rho^{(0)}_x$ and $\Delta^{(0)}$ are the radius and center of the reference path in Fig. \ref{Fig2} along the $\Delta$ coordinate. The results are shown in Fig. \ref{Fig3}, where we observe that despite large variations in the path the $\alpha$ enantiosensitivity of the asymmetric switch effect is maintained (see also SI \cite{SuppNotes}).

Crucially, even in the case of perfect control of the experimental parameters, the position of the EPs depends on the molecular orientation in the laboratory frame via the projection of the molecular dipoles on the polarization axis of the fields driving the corresponding transitions $\mathbf{d}_{ij}\cdot\mathbf{F}_\alpha$. Effectively, accounting for all molecular orientations results in a distribution of EPs over the parameter space. While we observe that the distributions of EPs of the two enantiomers are mirror versions of each other when driven by a synthetic chiral field $\left[\mathbf{F}^*_{1}\times\mathbf{F}_{2}\right]\cdot\mathbf{F}_{3}\neq0$ (see Fig. 3 of the SI \cite{SuppNotes}), the asymmetric switch is essentially averaged out when looking at the orientational averaged population in the adiabatic states of the enantiomers, as quantified by the parameter $\langle\alpha\rangle\simeq0.47$ for the right and $\langle\alpha\rangle\simeq0.41$ for the left enantiomer, where the brackets indicates orientational averaging (see Appendix A). Nonetheless, about 38\% of orientations of the right enantiomer exhibit a negative $\alpha$, indicating a robust topological population transfer for those orientations whose EP lie inside the chosen path. For the left enantiomer, negative $\alpha$ values do occur for some orientations, but they are very close to zero ($\alpha\simeq-0.0025$), implying that the topological effect is effectively negligible. In particular, the enantiosensitivity can be fully preserved if the molecular gas is aligned and the fields' polarizations are chosen judiciously. For example, if the molecule is aligned such that $\mathbf{d}_{12}$ dipole points along the polarization axis of the $\mathbf{F}_{3}$ field and we choose the $\mathbf{F}_{1}$ and $\mathbf{F}_2$ fields to be co-rotating circularly polarized fields, a rotation of the bound-free molecular dipoles $\mathbf{d}_{iE}$ around the axis $\mathbf{F}_3$ by an angle $\beta$ will lead to the transition matrix elements $M_{iE}=\mathbf{d}_{iE}\cdot\mathbf{F}_i$ to pick up an equal phase $M_{iE}(\beta)=M_{iE}(0)\exp(\pm\text{i}\beta)$. Consequently, these phases will cancel out in the resulting three-photon cyclic matrix element $\Omega_{123}(\beta)$ (see Eq. [\ref{eq:O123}]), leaving the position of the EPs in parameter space unaffected, and hence the corresponding enantiosensitive asymmetric switch effect. We refer the reader to the SI \cite{SuppNotes} for a more in-depth discussion on the effect of molecular orientation and additional results.

\begin{center}
\begin{table}[]
\begin{tabular}{|l|c|c|c|c|}

\multicolumn{1}{c}{} & \multicolumn{2}{c}{R} & \multicolumn{2}{c}{L} \\
\cline{2-5}
\multicolumn{1}{c|}{} & $\phi_+(0)$ & $\phi_-(0)$ & $\phi_+(0)$ & $\phi_-(0)$ \\
\hline
$\phi_+(T_{\mathrm{loop}})\ \circlearrowleft$  & 0.40\% & 2.24\% & 0.76\% & 2.98\% \\
$\phi_-(T_{\mathrm{loop}})\ \circlearrowleft$  & 1.12\% & 4.17\% & 2.30\% & 9.00\% \\
$\phi_+(T_{\mathrm{loop}})\ \circlearrowright$ & 4.17\% & 2.24\% & 0.76\% & 2.30\% \\
$\phi_-(T_{\mathrm{loop}})\ \circlearrowright$ & 1.12\% & 0.40\% & 2.98\% & 9.01\% \\
\hline
\end{tabular}
\caption{Population left in the adiabatic states in the two enantiomers after encirclement of the EP of the right enantiomer. Each entry corresponds to the population left in a given adiabatic state $\phi_\pm$, for clockwise $\circlearrowright$ or counter-clockwise $\circlearrowleft$ encirclement, where each column corresponds to different initial conditions. The first two columns from the left correspond to the right enantiomers, the last two columns to the left enantiomer.}
\label{tab:respop}
\end{table}
\end{center}

\section{Enantiosensitive stabilization of quasistationary states}
In the linear regime, chiral light-matter coupling requires interaction with the magnetic field. However, since the magnetic field couples to matter weakly, the enantiosensitive response is typically several orders of magnitude smaller than the absorption of light at the same frequency. The strength of the chiral effects is dictated by the ratio of electric and magnetic dipoles $\epsilon=\Omega_{m}/\Omega_d\simeq 1/c\ll1$, where $\Omega_{m}=-\mathbf{m}\cdot\mathbf{B}$ and $\Omega_{d}=-\mathbf{d}\cdot\mathbf{E}$ and $c$ is the speed of light. 

Yet, what happens when the electric and magnetic transitions drive population to a metastable state? Here we show that in the vicinity of such a resonance, despite the weak chiral light-matter interaction, a large enhancement of enantiosensitive effects can be observed.

\begin{figure*}
\begin{center}
\includegraphics[width=17cm, keepaspectratio=true]{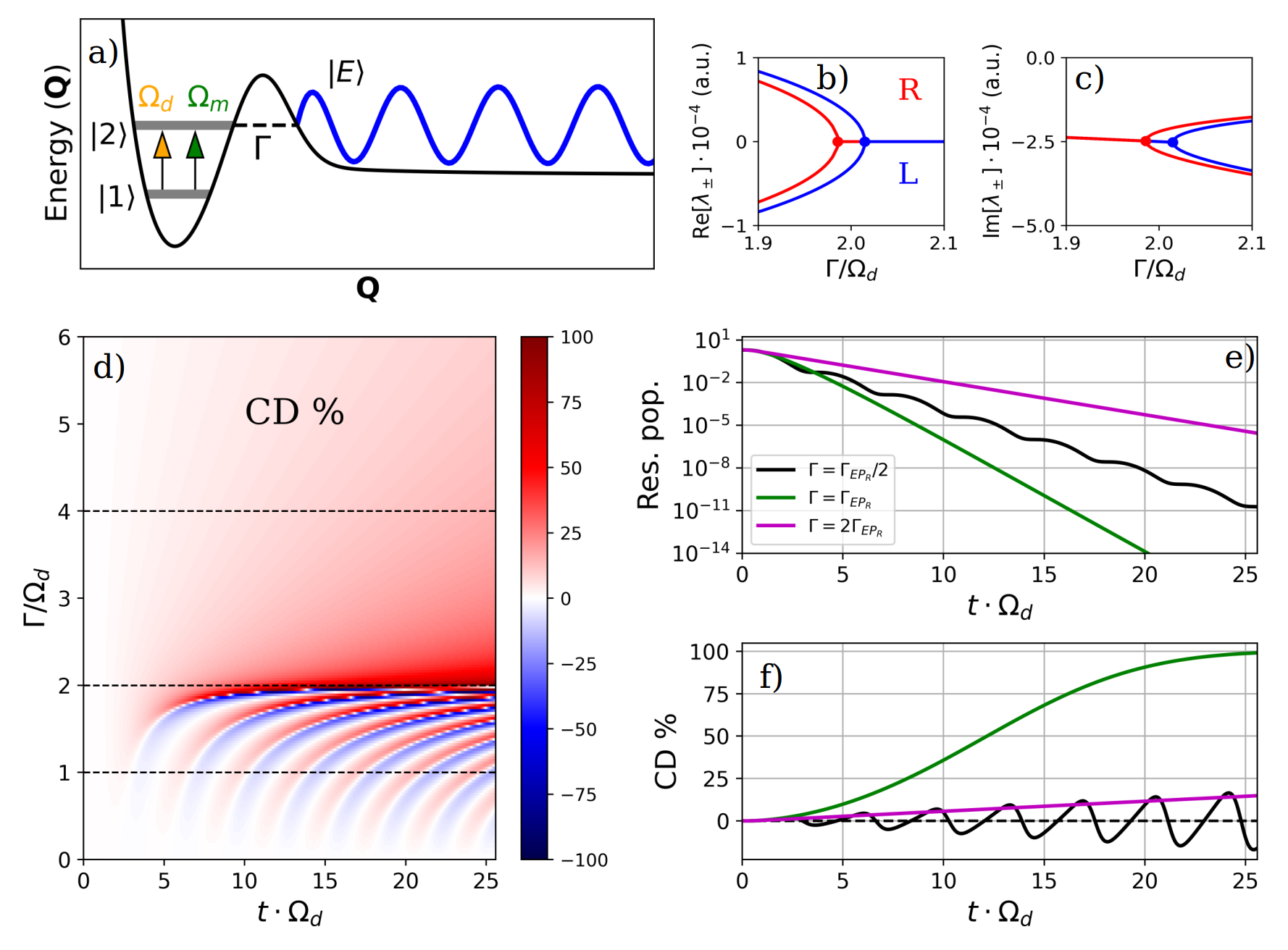}
\caption{\textbf{Amplification of weak chiral coupling via non-Hermitian effects.} \textbf{(a)} Two levels $|1\rangle$ and $|2\rangle$ of a chiral molecule are coupled by a circularly-polarized field via electric-dipole and magnetic-dipole transitions (orange and green arrows). The upper level is coupled via tunneling through a barrier to the continuum $|E\rangle$ with a decay rate $\Gamma$. $\mathbf{Q}$ represents a general molecular coordinate. \textbf{(b,c)} Real (\textbf{b}) and imaginary (\textbf{c}) parts of the eigenvalues $\lambda_{\pm}$ in Eq. [\ref{eq:eig0}] with respect to the scaled decay rate $\Gamma/\Omega_d$. The red and blue colors correspond to right and left enantiomer respectively, and their respective EPs are indicated by colored circles. \textbf{(d)} Time-dependent chiral dichroism $\text{CD}(t)$ for increasing scaled decay rate $\Gamma/\Omega_d$ obtained from the solution of the Time-Dependent Schr{\"o}dinger Equation (TDSE) with the Hamiltonian in Eq. [\ref{eq:Ham0}] at zero detuning $\Delta=0$. The time is expressed in dimensionless units $t\cdot\Omega_d$. \textbf{(e}) Total residual population in the two enantiomers $P_+(t)+P_-(t)$ for selected values of the scaled decay rate $\Gamma/d_{12}$. \textbf{(f)} Time-dependent chiral dichroism for selected values of the scaled decay rate as in panel (d). The horizontal dashed line indicates $\text{CD}(t)=0$. }
\label{Fig4}
\end{center}
\end{figure*}

The required resonance conditions emerge naturally in a broad range of systems \cite{Piancastelli:1999,moiseyev_2011}. Consider for example two vibrational states of a chiral molecule that tunnel through a barrier into the dissociation continuum, as in the case of a shape resonance shown in Fig. \ref{Fig4}a. If the two discrete levels are coupled by a light field via a magnetic $\Omega_m$ and dipole $\Omega_d$ transition, the Hamiltonian describing the evolution of the open two-level system in the Rotating-Wave Approximation (RWA) is given by
\begin{equation}\label{eq:Ham0} H=\begin{bmatrix}0 & \Omega_m+\Omega_d \\ \Omega_m^*+\Omega_d^* & \Delta-\text{i}\Gamma\end{bmatrix},\end{equation}
where $\Gamma$ is the tunneling rate and $\Delta=E_2-E_1-\omega_L$ is the detuning from the transition frequency, where $\omega_L$ is the driving field frequency.

The eigenvalues of the Hamiltonian are given by
\begin{equation}\label{eq:eig0}\lambda_{\pm}=\frac{\Delta-\text{i}\Gamma\pm\sqrt{\delta}}{2},\end{equation}
with the discriminant $\delta$
\begin{eqnarray}\delta&=&(\Delta-\text{i}\Gamma)^2+4|\Omega_m+\Omega_d|^2\end{eqnarray}
becoming enantiosensitive due to the factor $|\Omega_m+\Omega_d|^2$. Indeed, the squared absolute value of the sum of two Rabi frequencies depends on the sign of the cross term $\Omega_m\cdot\Omega_d$ -- the typical parameter quantifying enantiosensitivity in optical rotation or absorption circular dichroism. After orientational averaging this parameter represents a pro\-duct of light pseudoscalar known as optical chirality \cite{Tang:2010aa} and molecular pseudoscalar $\mathbf{d}\cdot\mathbf{m}$. Thus if  $\mathbf{d}\cdot\mathbf{m}\neq 0$ the term $\Omega_m\cdot\Omega_d$ is enantiosensitive, flipping sign with the molecular handedness. Thus, the EPs of two enantiomers in the parameter space defined by the detuning $\Delta$ and decay rate $\Gamma$ are again separated:
\begin{equation}\left(\Delta^{EP_{R}},\Gamma^{EP_{R}}\right)=\left(0,\pm2|\Omega_m+\Omega_d|\right),\end{equation}
\begin{equation}\left(\Delta^{EP_{L}},\Gamma^{EP_{L}}\right)=\left(0,\pm2|\Omega_m-\Omega_d|\right).\end{equation}
The real and imaginary parts of the eigenvalues $\lambda_{\pm}$ with respect to the scaled decay rate $\Gamma/\Omega_d$, for a typical ratio of electric and magnetic Rabi frequencies of $\Omega_d/\Omega_m=c$, exhibit a characteristic behavior in the vicinity of an EP (see Figs.\ref{Fig4}b,c, where red (blue) lines correspond to the right (left) enantiomers. For decay rates $\Gamma<\Gamma^{EP}$ the two adiabatic solutions have equal imaginary parts (equal decay rates) but opposite real parts (resonant energy), leading to a time-dependent decaying bound population oscillating at frequency $2\pi/|\mathrm{Re}(\lambda_+)-\mathrm{Re}(\lambda_-)|$ (see the black solid line in Fig. \ref{Fig4}e). For $\Gamma>\Gamma^{EP}$ the real part of the solutions coalesce and the imaginary part branches, with one adiabatic solution becoming stabilized as its imaginary part becomes increasingly small. Correspondingly, the decay of the bound population counter-intuitively slows down for increasing $\Gamma$.

This lifetime branching behavior in non-Hermitian systems is at the core of atomic and molecular phenomena like interference stabilization \cite{Fedorov:1988aa,Fedorov:2011aa,Eckstein:2016aa,Mies:1968aa} as well as prompt and delayed dissociation in molecules \cite{Remacle:1996aa,Remacle:1996ab,Reitsma:2019aa}. In the language of optics and photonics \cite{Ozdemir:2019aa}, the EP separates the $\mathcal{PT}$-symmetric region $(\Gamma<\Gamma^{EP})$ from the $\mathcal{PT}$-broken one $(\Gamma>\Gamma^{EP})$.

This behavior is seen in both enantiomers, but the slight shift in the positions of the EPs leads to dramatic differences. To gauge their magnitude, let us explore the splitting of real and imaginary energies at the vicinity of exceptional points $4\Omega_d^2-\Gamma^2=0$ at the resonance $\Delta=0$:
\begin{equation}
\label{eq:zeroDelta}
\lambda_{\pm}=\frac{-\text{i}\Gamma\pm\sqrt{8\epsilon\Omega_d^2}}{2}.
\end{equation}
Since $\epsilon=\Omega_m/\Omega_d$ has opposite sign in left and right enantiomers, we obtain for $\epsilon>0$:
\begin{equation}
\label{eq:zeroDelta}
\lambda_{\pm}=\frac{\Gamma}{2} \left[ \text{-i}\pm \sqrt{2|\epsilon|}\right]
\end{equation}
while for $\epsilon<0$:
\begin{equation}
\label{eq:zeroDelta}
\lambda_{\pm}=\frac{i\Gamma}{2} \left [ -1\pm \sqrt{2|\epsilon|}\right]
\end{equation}
Thus, the enantiomer corresponding to negative $\epsilon$ decays faster by a factor $\frac{1}{2}\sqrt{|\epsilon|}$. The square root scaling features an order of magnitude enchancement of the  enantiosensitive  response  relative to the standard regime, where it scales as $\epsilon$.  

The enantiosensitivity of this stabilization phenomenon is clearly seen in the simulations when looking at the circular dichroism in the residual population $\text{CD}=(P_R-P_L)/(P_R+P_L)$, see Fig. \ref{Fig4}, where we fix $\Omega_d=2.5\cdot10^{-4}$ a.u. and $\Omega_m=\Omega_d/c$ (see Appendix B for further details on the simulations) and express the time in dimensionless units $t\cdot\Omega_d$. For $\Gamma<\text{min}(\Gamma^{EP_R},\Gamma^{EP_L})$, both enantiomers are in the $\mathcal{PT}$-symmetric phase and the bound population oscillates between the ground and excited state with a lifetime of $\Gamma$. The slight difference in the oscillation periods of the two enantiomers $T_{R/L}=2\pi/\mathrm{Re}[\lambda^{R/L}_+-\lambda^{R/L}_-]$ leads to an oscillating $\text{CD}$ at the average period $T=(T_R+T_L)/2$ (black line in Fig. \ref{Fig4}f). While the amplitudes of the $\text{CD}$ oscillations can be quite large, they effectively average to zero over the beating period $T$ (dashed black line in Fig. \ref{Fig4}f). As $\Gamma$ approaches the EPs of the two enantiomers, the oscillation periods become increasingly larger as the real part of the adiabatic energies approach each other.
When $\Gamma<\Gamma^{EP_L}$ and $\Gamma>\Gamma^{EP_R}$, the left enantiomer is in the $\mathcal{PT}$-symmetric phase and the right enantiomer is in the $\mathcal{PT}$-broken one. In this region the different topological phases of the enantiomers lead to a $\text{CD}$ that saturates quickly to the theoretical maximum of $100\%$ after $t\cdot\Omega_d\simeq20$ (solid green line in Fig. \ref{Fig4}f). Still, the residual population in the enantiomers is largely depleted. For even larger $\Gamma$ (solid magenta line in Fig. \ref{Fig4}f), both enantiomers are in the $\mathcal{PT}$-broken phase and are described by bi-exponential behavior. Because of the stabilization effect, the lifetime of the bound population increases for increasing $\Gamma$. Here, small differences in the lifetimes $\tau_{R/L}=1/\text{Im}[\lambda^{R/L}_+]$ accumulate in time, leading to a $\text{CD}$ of $\simeq20\%$ after $t\cdot\Omega_d\simeq 25$ with a residual bound population of $\simeq 10^{-4}$.

Our results thus show that weak chiro-optical effects relying on magnetic dipole interactions can nonetheless lead to huge chiral dichroism effects owing to the enantiosensitive stabilization effect. The enhancement of these weak effects is most prominent when one of the enantiomer is in its $\mathcal{PT}$-broken phase and the other one is in the $\mathcal{PT}$-symmetric one. Yet these effects do not vanish when both enantiomers are in their $\mathcal{PT}$-broken phase, owing to the enantiosensitive stabilization effect. Effectively for $\Gamma\gg2\Omega_d$ the trade-off between the chiral dichroism $\text{CD}\simeq\tanh(2T/(\tau_R-\tau_L))$ and the amount of population left in the bound states $w_{\mathrm{res}}\simeq\exp(-T/\tau_{R/L})$ is dictated by the duration of the interaction $T$. It is therefore possible to fine tune and control the trade-off between enantiosensitivity and residual bound population by changing the duration of the laser pulse driving the transition. In an experimental realization of this idea, an ideal experimental observable could be the ionization yield of the molecular sample using standard time of flight mass spectrometry. We also note that in a realistic molecule additional decay channels will be present, which would require an expansion of the basis of molecular states considered. Nonetheless, even in this case stabilization effects in molecules are predicted \cite{Remacle:1996aa,Remacle:1996ab,Reitsma:2019aa}, which we expect to hold in chiral samples. Similarly to the asymmetric switch effect discussed earlier different molecular orientations can result in a shift of the position of EP, but molecular alignment is sufficient to retain the enantiosensitivity.

\section{Chiral optical fibers}
Optical fibers are widely used as sensors, combining the high sensitivity, versatility, and rapid detection typical of optical methods with in-situ measurement capability. Optical fibers are also one of the preferred experimental platforms to observe non-Hermitian effects, given that their gain/loss profile can be engineered \cite{Doppler:2016aa,Ozdemir:2019aa,Feng:2017aa,Parto:2020aa}. Owing to the high sensitivity of a non-Hermitian system tuned to an EP to external perturbations, EP-based sensors are promising candidates as highly responsive next-generation sensors \cite{Chen:2017aa,Lai:2019aa,Ruan:2025aa,Hodaei:2017aa,Hokmabadi:2019aa}. Here, we show how one can exploit the nature of EPs as critical points separating the $\mathcal{PT}$-symmetric and $\mathcal{PT}$-broken regions of a non-Hermitian fiber to detect molecular chirality in solution with high sensitivity.
\begin{figure*}
\begin{center}
\includegraphics[width=18cm, keepaspectratio=true]{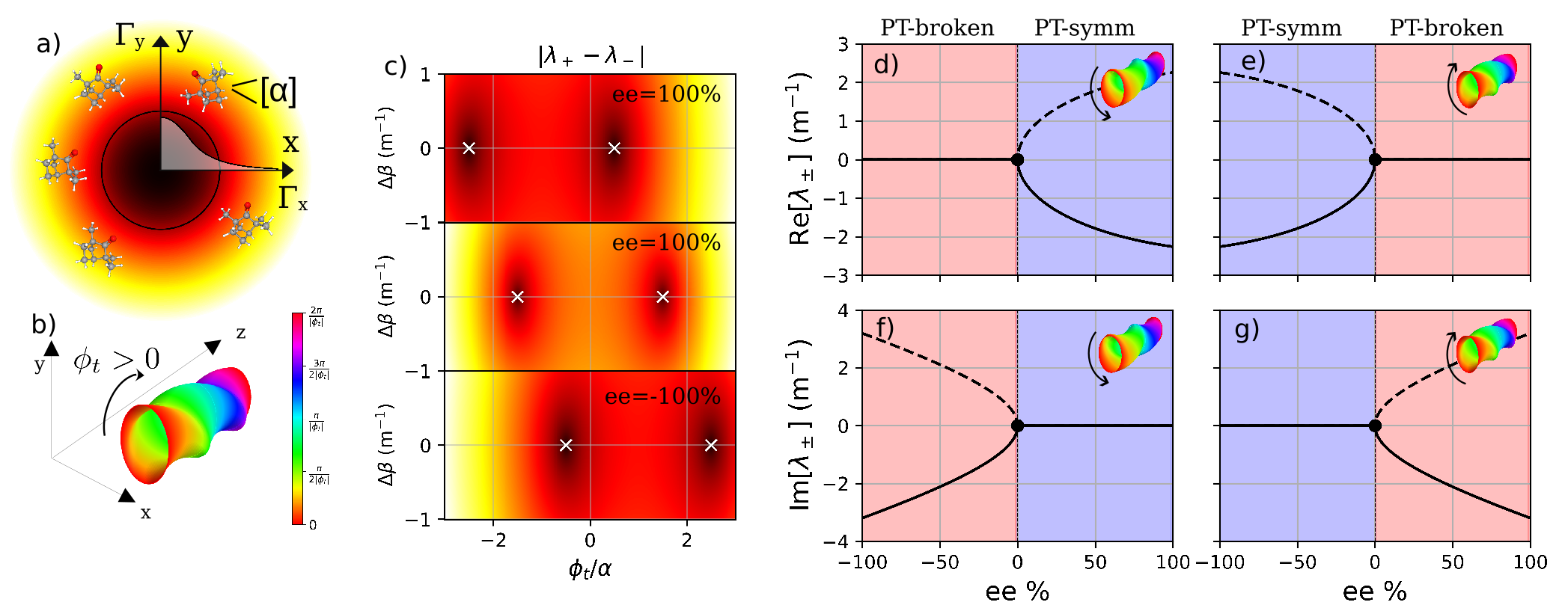}
\caption{\textbf{(a)} A single mode non-Hermitian optical fiber interacting with a solution of chiral molecules via its evanescent modes. \textbf{(b)} The fiber is twsited along its propagation axis $z$ with a torsion rate $\phi_t$, where positive $\phi_t$ indicates clockwise twist. \textbf{(c)}~ Absolute value of the difference between the eigenvalues of the fiber for enantiomeric excess $ee=100\%$ (top panel), $ee=0\%$ (middle panel) and $ee=-100\%$ (bottom panel). The white crosses indicate the EP. \textbf{(d-f}) Real (\textbf{d}) and imaginary (\textbf{f}) part of the eigenmodes of a fiber with negative torsion rate $\phi_t$. Red and blue areas indicate respectively the $\mathcal{PT}$-broken and $\mathcal{PT}$-symmetric regions. \textbf{(e,g)} Same as in (\textbf{f,h}) for a fiber with positive torsion rate $\phi_t$.}
\label{Fig5}
\end{center}
\end{figure*}
\par We consider a single-mode birefringent optical fiber, where the strong light confinement leads to an evanescent component of the propagating modes that interacts with a solution of randomly oriented chiral molecules in which the fiber is immersed (see Fig. \ref{Fig5}a). We describe the gain/loss profile of the fiber via the rates $\Gamma_x$ and $\Gamma_y$ along the two laboratory axes $\hat{x}$ and $\hat{y}$. Negative (positive) $\Gamma_i$ corresponds to loss (gain), and we denote by $\Delta\Gamma=\Gamma_x-\Gamma_y$ the relative gain/loss. Similarly, $\Delta\beta=\beta_x-\beta_y$ describes the birefringence of the fiber along the same two axes. Twisting the fiber along its propagation axis $z$ (Fig. \ref{Fig5}b) makes it chiral, with its handedness described by the torsion rate, a pseudoscalar quantity given by
\begin{equation}\label{eq:torsion}\phi_t=\mathrm{Im}\left[\left(\mathbf{\Psi}(z)\times\frac{d\mathbf{\Psi}(z)}{dz}\right)\cdot\mathbf{z}\right],\end{equation}
where positive (negative) $\phi_t$ corresponds to a clockwise (counter-clockwise) twist along the $z$ axis.
Here $\mathbf{\Psi}(z)=R_z(\phi_t z)\mathbf{\Psi}$, $R_z$ is the rotation matrix about the $z$-axis and $\mathbf{\Psi}=\left(\Psi_-,\Psi_+\right)/\sqrt{2}$ is the propagating mode in the laboratory frame, where $\Psi_\pm=(\Psi_x\mp\text{i}\Psi_y)/\sqrt{2}$ are the RCP and LCP modes.

The fiber twist introduces a phase shift between the circularly-polarized components that is directly proportional to the torsion rate $\phi_t$. Hence, a linearly polarized mode experiences optical rotation as it propagates in the twisted fiber. Similarly, the interaction with the surrounding solution of chiral molecules will also lead to optical rotation, now directly proportional to the enantiomeric excess $ee=C_R-C_L$, where $C_R$ ($C_L$) is the concentration of right (left) molecules in the solution. 

How does the enantiomeric excess of the solution affect the light propagating through this fiber? Is it possible to detect the enantiomeric excess by twisting the fiber so to match the optical rotation induced by the molecules?
\begin{figure*}
\begin{center}
\includegraphics[width=18cm, keepaspectratio=true]{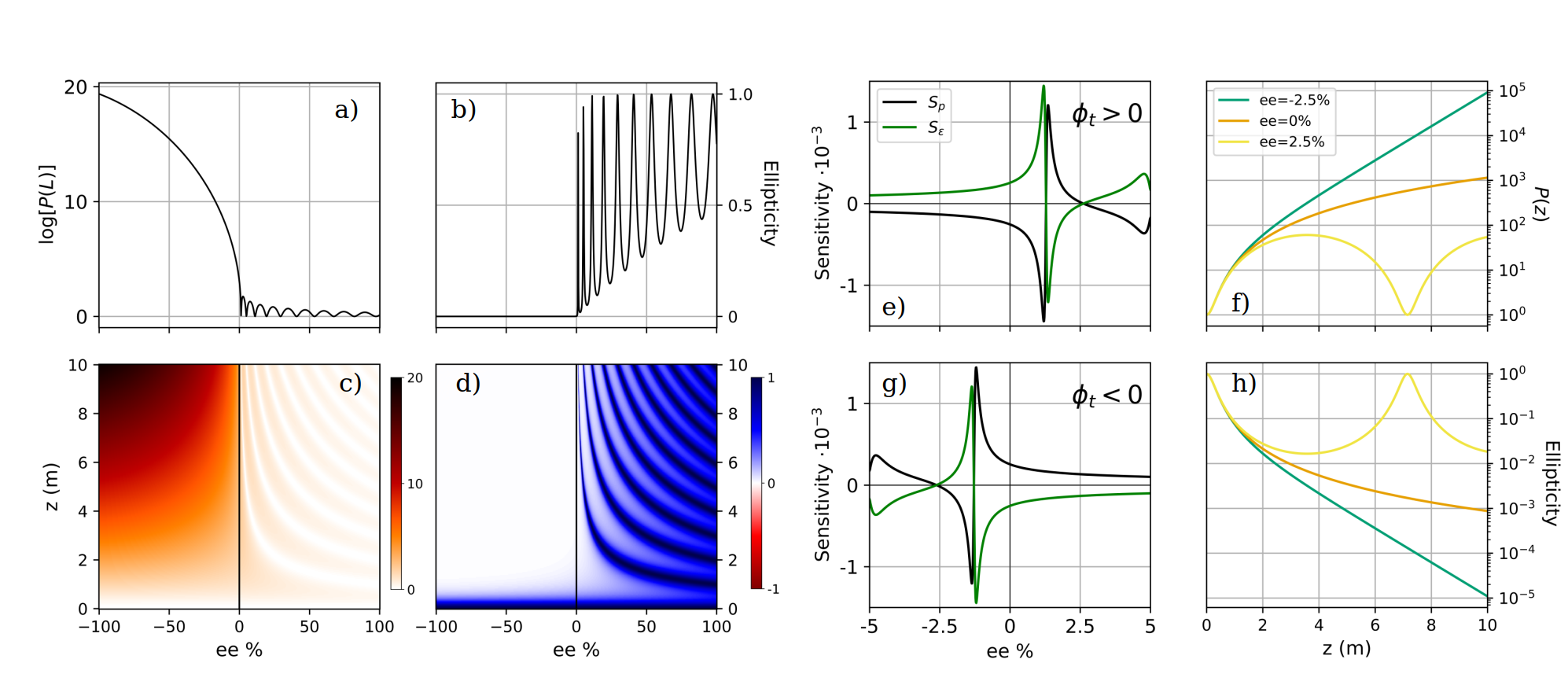}
\caption{Numerical results for a single mode PMMA fiber operating at $\lambda=589$ nm and interacting with a solution of fenchone molecules. \textbf{(a,b)} Power $P(z)$ in logarithmic scale (\textbf{a}) and ellipticity $\xi(z)$ (\textbf{b}) of the modes at the end of a fiber with length $L=10$ m and torsion rate $\phi_t=2.39$ m$^{-1}$, as a function of the enantiomeric excess of the solution $ee$.\textbf{(c,d)} Power in logarithmic scale (\textbf{c}) and ellipticity (\textbf{d}) in the modes as a function of $ee$ and propagation distance $z$. \textbf{(e,g)} Sensitivity of the fiber response to variations in $ee$, for \textbf{e} $\phi_t=2.39$ m$^{-1}$ and \textbf{g} $\phi_t=-2.39$ m$^{-1}$. Black and green lines correspond respectively to the sensitivity calculated from the power $P(z)$ and ellipticity $\xi(z)$ at the end of the fiber. \textbf{(f,h)} Power (\textbf{f}) and ellipticity (\textbf{h}) of the modes as a function of propagation distance $z$ for enantiomeric excess $ee=-2.5\%$ (cyan lines), $ee=0\%$ (orange lines) and $ee=2.5\%$ (yellow lines) for a fiber with torsion rate $\phi_t=2.39$ m$^{-1}$.}
\label{Fig6}
\end{center}
\end{figure*}

As derived in the SI \cite{SuppNotes}, the circularly-polarized modes in the frame co-rotating with the fiber follow a  non-Hermitian evolution $\text{i}\partial_z\mathbf{\Psi}(z)=H\mathbf{\Psi}(z)$, where $H=H_0+H_{ee}$ and
\begin{eqnarray}\label{eq:HamFIB}H_0&=&\begin{bmatrix}-\phi_t & \Delta\beta-\text{i}\Delta\Gamma \\
\Delta\beta-\text{i}\Delta\Gamma & \phi_t\end{bmatrix},\\
H_{ee}&=&ee\begin{bmatrix} \alpha & 0 \\ 0 & -\alpha\end{bmatrix}.\nonumber\end{eqnarray}
Here $H_0$ is the Hamiltonian of the twisted non-Hermitian fiber and $H_{ee}$ accounts for the interaction with the chiral medium with enantiomeric excess $ee$, where $\alpha$ is the specific optical rotation of the chiral molecules in the solution. The eigenvalues of the matrix $H$ are the complex propagation constants $\beta$ of the eigenmodes along the $z$ axis: their real part is the phase constant and the imaginary part is the attenuation or gain coefficient. Note that in Eq. [\ref{eq:HamFIB}] we have removed the common gain/loss factor of the two modes $\Gamma=\Gamma_x+\Gamma_y$.

The position of the enantiosensitive EPs in the parameter space defined by the torsion $\phi_t$ and the birefringence $\Delta\beta$ for a given $\Delta\Gamma$ are (see SI \cite{SuppNotes})
\begin{eqnarray}
(\phi^{EP_1}_t,\Delta\beta^{EP_1})&=&(ee\,\alpha-\Delta\Gamma,0),\\
(\phi^{EP_2}_t,\Delta\beta^{EP_2})&=&(ee\,\alpha+\Delta\Gamma,0).
\end{eqnarray}
Hence, the interaction between the fiber's modes and the molecules leads to enantiosensitive EPs whose position depends on the chirality (enantiomeric content) of the solution, with EPs of solutions with opposite handedness separated by $\Delta\Gamma$. In Fig. \ref{Fig5}c we show the absolute value of the difference between the eigenmodes $|\lambda_+-\lambda_-|$ for $ee=-100\%,\,0\%$ and $100\%$. 

Crucially, if the torsion of the fiber is fixed and the enantiomeric excess is varied, we observe a strikingly different behavior depending on the direction of the twist. This is shown in Figs. \ref{Fig5}e-h), where we fix $\Delta\Gamma=1.5\alpha$ and consider clockwise and counter-clockwise twisted fibers tuned to the EP corresponding to the racemic solution $ee=0\%$, $\phi_t=\pm\Delta\Gamma$. In Figs. \ref{Fig5}e,g the fiber has a negative (counter-clockwise) twist $\phi_t=-\Delta\Gamma$ (assuming $\Delta\Gamma>0$), while in Figs. \ref{Fig5}f,h the twist is positive (clockwise) $\phi_t=\Delta\Gamma$. In both cases $\Delta\beta=0$. When the mixture is racemic $ee=0\%$, the fiber is at an EP for both $\phi_t>0$ and $\phi_t<0$, as expected. If there is an excess of left-handed molecules in the solution ($ee<0\%$), the fiber with negative torsion rate is in the $\mathcal{PT}$-broken regime: the eigenmodes propagate with same velocity (same real part $\mathrm{Re}[\lambda_\pm]$, see Fig. \ref{Fig5}e) but with different gain/attenuation rate (different imaginary part $\mathrm{Im}[\lambda_\pm]$, see Fig. \ref{Fig5}g). Strikingly, if the fiber is twisted in the clockwise sense ($\phi_t$, Figs. \ref{Fig5}f,h), the modes propagate in the $\mathcal{PT}$-symmetric regime: the gain/attenuation rate is the same (see Fig. \ref{Fig5}h) but the velocity at which they propagate differs (Fig. \ref{Fig5}f).

Hence, topologically-distinct behavior is observed depending on the relative handedness of the fiber (the torsion rate) and of the solution (the enantiomeric content). Notably, identical behavior can be seen at \textit{any} enantiomeric excess $ee$, provided that the fiber's torsion is at the corresponding EPs $\phi_t=ee\pm\Delta\Gamma$. 

We now show how this topological difference affects the modes propagating in the fiber. We couple RCP light with wavelength $\lambda=589$ nm into a Poly(methyl methacrylate) (PMMA) fiber with radius $0.5\mu$m, ensuring single-mode operation. The evanescent tail of the guided modes interacts with a solution of fenchone molecules, whose refractive index and optical rotation is calculated at the B3LYP/aug-cc-pVDZ level using Gaussian16 \cite{g16} (see Appendix C for more details). Figs. \ref{Fig6}a-d show the results for a fiber with positive twist $\phi_t=|\Delta\Gamma|=2.39$ m$^{-1}$, where Figs. \ref{Fig6}a,c show the power $P(z)=\sum_{i=\pm}|\Psi_i(z)|^2$ (in logarithmic scale) and Figs. \ref{Fig6}b,d show the ellipticity of the light $\xi(z)=(|\Psi_+(z)|^2-|\Psi_-(z)|^2)/P(z)$. Figs. \ref{Fig6}a-b show the intensity and ellipticity at the end of the fiber $L=10$ m. Note that since we removed the common gain/loss factor $\Gamma=\Gamma_x+\Gamma_y$ in Eq. [\ref{eq:HamFIB}] the power $P(z)=\sum_{i=\pm}|\Psi_i(z)|^2$ should be understood as the relative gain or loss.

When $ee<0$ the fiber is in the $\mathcal{PT}$-broken regime and the modes propagate in phase ($\mathrm{Re}[\lambda_+]=\mathrm{Re}[\lambda_-]$) but with different gain/attenuation constants ($\mathrm{Im}[\lambda_+]\neq\mathrm{Im}[\lambda_-]$). Correspondingly, we observe an exponential increase of the relative gain that has the typical square root scaling of EPs with respect to the enantiomeric excess $I(z)\propto\exp(\sqrt{ee} \cdot z)$. Moreover, the modes become quickly linearly-polarized during their propagation. For $ee>0$, the fiber is instead in the $\mathcal{PT}$-symmetric regime and the modes propagate with same gain/attenuation constants ($\mathrm{Im}[\lambda_+]=\mathrm{Im}[\lambda_-]$) but out of phase ($\mathrm{Re}[\lambda_+]\neq\mathrm{Re}[\lambda_-]$). Correspondingly, we observe beatings in the total power and ellipticity of the modes with a period $T \propto\sqrt{ee}$, where the maximum ellipticity corresponds to the minimum in relative power $P(z)$. The racemic case marks the transition between these two regimes at the enantiosensitive EP.

The topological transition associated with an excess of right- or left-handed molecules makes this fiber-based sensor exceptionally sensitive to small variations in enantiomeric excess, on par with other topological methods that have been proposed recently \cite{Mayer:2024aa,Schwennicke:2022aa}. To characterize the response of the fiber to small variations in $ee$, we define its sensitivity as $R(ee)=\frac{1}{|S(ee)|}\frac{dS(ee)}{dee}$, where $S(ee)$ is either the power $P(L)$ or ellipticity $\xi(L)$ at the end of the fiber $z=L$. Figs. \ref{Fig6}e show the sensitivity of the fiber for a positive $\phi_t=2.39$ m$^{-1}$ (Fig. \ref{Fig6}e) or negative $\phi_t=-2.39$ m$^{-1}$ (Fig. \ref{Fig6}g) torsion, where black (green) lines correspond to the sensitivity associated to the power (ellipticity). The resonant behavior at $|ee|\simeq1.25\%$ is a consequence of the oscillations in the intensity $P(z)$ or ellipticity $\xi(z)$ in the $\mathcal{PT}$-symmetric regime. At $ee=0\%$, the sensitivity is still sufficient to clearly distinguish small variations in enantiomeric excess, as clearly seen also in Figs. \ref{Fig6}f-h, where we show the power and ellipticity in the modes along for $ee=-2.5\%,\,0\%$ and $2.5\%$ for a fiber with positive torsion $\phi_t=2.39$ m$^{-1}$. The sensitivity for a fiber with negative torsion $\phi_t=-2.39$ m$^{-1}$ is shown in Fig. \ref{Fig6}g.

Our results thus show that a chiral non-Hermitian fiber can be employed as a highly sensitive sensor of molecular chirality, where small variations of enantiomeric excess are mapped onto topologically distinct modes of operation of the sensor. We stress once again that this topological transition can be engineered to detect small variations in the concentration of enantiomers in a solution $\delta=ee-ee^{(0)}$ around any given enantiomeric excess $ee^{(0)}$, as long as the torsion is tuned to a corresponding EP $\phi_t=ee^{(0)}\alpha\pm\Delta\Gamma$. This is in stark contrast with other topological methods, where the high sensitivity associated with topological transition is observed only near $ee=0\%$ \cite{Mayer:2024aa}. We also note that the rotation angle of the polarization ellipse is also a valid observable, where similar behavior can be seen (not shown here).

\section{Outlook and conclusions}

In this work we have introduced and discussed the new concept of enantiosensitive exceptional points in chiral molecular non-Hermitian systems. Using intuitive and minimal models describing the interaction between chiral light and matter, we have shown how by appropriately tuning the system we can separate the EPs associated to molecular enantiomers of opposite handedness in the parameter space of the non-Hermitian system.

Using a three-color field coupling a chiral molecule to the continuum, we have demonstrated enantioselective topological population transfer by enclosing the enantiosensitive EP of only one of the molecular enantiomers, while its mirror twin's EP remains outside of the enclosing path. When compared to other -- Hermitian -- enantiosensitive population transfer mechanisms, such as three-wave mixing \cite{Eibenberger:2017aa,Patterson:2013aa} or coherently controlled passage \cite{Kral:2007aa}, the non-Hermitian protocol presented here offers stability and efficiency against common experimental noise owing to its topological nature. While random molecular orientations can reduce the efficiency of the topological population transfer, molecular alignment along one axis can be sufficient to recover the effects, as shown in more details in the SI \cite{SuppNotes}. We also note that orientation of small chiral molecules is well within experimental reach \cite{Wade:2022aa}.

We have then shown preferential ionization and/or dissociation of chiral molecules in the proximity of enantiosensitive EPs, resulting in large chiral dichroism despite linear regime, where chiroptical response is several orders of magnitude smaller than light absorption at the same frequency. This result is particularly interesting for the spatial separation of enantiomers and the enantiomeric enrichment of racemic mixtures. We report an order of magnitude enchancement of a signal compared to  other photochemical methods like asymmetric photodestruction \cite{Kang:2022aa,Feringa:1999aa}, and can be further enhanced by tuning the enantiosensitive interference between the light-driven transitions using synthetic chiral light \cite{Ayuso:2019aa}. While the use of a two-level model in a realistic case of nuclear or electronic motion of a molecule interacting with an external field is a simplification that is well justified only for comparably cold molecules, one can use e.g. supersonic expansion of gas jets to cool down the sample. Moreover our results can be generalized to a larger basis of molecular states using the non-Hermitian formalism 
\cite{Mies:1968aa,Fedorov:2011aa}. In general if the number $N$ of bound states is larger than the number $K$ of decay channels, the lifetime branching phenomena explored in this paper results in $N-K$ states becoming stabilized for increasing average width $\langle\Gamma\rangle$ of the resonances, as seen in Refs. \cite{Remacle:1996aa,Remacle:1996ab,Reitsma:2019aa}. Once chiral interactions are taken into account, one would expect an enantiosensitive split of the critical $\langle\Gamma^{EP}\rangle$ value for the two enantiomers to occur, leading to enantiosensitive losses toward the ionization or dissociation continuum. As mentioned in the text, the trade-off between chiral dichroism and residual bound population is decided by the duration of the interaction, which can be easily controlled in the experiment via the use of pulsed laser fields. The few-level model discussed here could also be extended to other resonant phenomena like autoionization or laser-induced continuum structures \cite{Knight:1990aa}, where ultrafast techniques like attosecond transient absorption \cite{Wu:2016aa,Drescher:2016aa,Drescher:2025aa} could unveil the underlying ultrafast chiral dynamics. 

Finally, we have shown how to use enantiosensitive EPs to develop new highly sensitive sensors of enantiomeric excess in chiral solutions. We envision this approach to pave the way for new chiral methods that find their application in clinical setups, where small imbalances in enantiomeric excess of biofluids samples can indicate diseases in humans \cite{Zheng:2023aa}. We have shown using realistic parameters for a PMMA fiber interacting with a solution of fenchone molecules that the EPs depend on the torsion rate of a non-Hermitian twisted fiber and the enantiomeric content of the solution. Taking advantage of the nature of EPs as critical points connecting the $\mathcal{PT}$-symmetric and $\mathcal{PT}$-broken regimes of a non-Hermitian system, we have shown how small variations in enantiomeric excess of a racemic solution lead to topologically different light propagation. This behavior is general, and the sensor can be made to be sensitive to small variations around any enantiomeric excess $ee^{(0)}$. In an experimental realization of our proposal, we note that typical values of the gain/loss profile $\Delta\Gamma$ are constrained by both the optical activity $\alpha$ of the chiral solution and the achievable torsion rates $\phi_t$ of the fiber. Since the EP condition requires $\phi_t=ee\cdot\alpha\pm\Delta\Gamma$, one expects $|\Delta\Gamma|$ to be comparable to the specific rotation $\alpha$, providing a lower bound $\Delta\Gamma\simeq 1$ m$^{-1}$ for typical molecular solutions. Experimentally accessible torsion rates in PMMA fibers can reach $10^{2}-10^{3}$ m$^{-1}$ \cite{Preizal:2024aa}, setting an upper bound on $\Delta\Gamma$. We also note that small drifts in torsion or birefringence can effectively detune the system from the exceptional point, shifting its location in parameter space. In practice, this implies that the sensor requires calibration to identify the operating point. However, since the eigenvalue splitting retains its characteristic square-root dependence on the distance from the EP, the enhanced sensitivity to variations in enantiomeric excess is preserved. In conclusion, the choice of the experimental platform where such ideas could be tested should be guided by the parameter ranges defined by the chiral solution, the material guiding the modes and the specific implementation of the gain/loss profile $\Delta\Gamma$. For example, the required interaction length described in the main text ($z \simeq 1$–$10$ m) can be achieved using a high-finesse resonator geometry based on a helically coiled optical fiber, where the effective torsion rate arises from the geometric phase accumulated by circularly polarized modes propagating along the helical waveguide \cite{Tomita:1986aa}. In this configuration, polarization-dependent attenuation can be introduced via evanescent coupling to an auxiliary waveguide placed in proximity to the coil, with the effective gain/loss contrast tunable through the separation between the two.

In conclusion, we introduced the concept of enantiosensitive exceptional points as a unifying framework for enantiosensitive techniques and non-Hermitian systems. Chirality is a ubiquitous phenomenon, spanning a wide range of temporal and length scales. Because realistic systems are inevitably open to their environment, they can be naturally described within a non-Hermitian formalism. The combination of these branches of research is very promising for both directions, as non-Hermiticity can provide enhancement of chiral phenomena, while chiral interactions open new avenues for the exploitation of EP-induced effects. Optical and photonic platforms \cite{Ozdemir:2019aa}, including cavity-based approaches \cite{Moiseyev:2023aa,Schafer:2023aa}, offer in this sense a promising direction to test these effects experimentally. Indeed it is on these platforms that EP-induced effects are routinely observed and technologically implemented \cite{Ozdemir:2019aa,Feng:2017aa,Xu:2016aa,Doppler:2016aa}. Solid-state topological platforms, including photonic Floquet topological insulators \cite{Rechtsman:2013aa}, are also a particularly suitable testbed for the ideas presented here.

\begin{acknowledgments}
Nicola Mayer gratefully acknowledges helpful discussions with Margarita Khokhlova and Emilio Pisanty. This project has received funding from the EU Horizon 2020 (grant agreement No 899794) and European Union (ERC, ULISSES, 101054696). This work was funded by UK Research and Innovation (UKRI) under the UK government’s Horizon Europe funding guarantee [grant number EP/Z001390/1]. Nimrod Moiseyev acknowledges Israel Science Foundation (ISF) grant No. 1757/24  for a partial support.
\end{acknowledgments}

\section*{Data availability statement}

The data that support the findings of this article are openly available at \href{https://doi.org/10.5281/zenodo.20488034}{https://doi.org/10.5281/zenodo.20488034}.

\appendix

\appendix

\section{Numerical details of the topological population transfer}

We encircle the EPs in the parameter space defined by the field coupling the two bound states via the one-photon dipole transition 
$(\Delta,F_3)$, where $\Delta=E_2-E_1-\omega_3$ and $F_3$ is the field strength. Such an encirclement can be achieved by considering chirped laser pulses \cite{Gilary:2012,Zdanska:2014aa,PhysRevA.88.010102}, where the chirp rate of the field at frequency $\omega_2$ has to be adjusted accordingly in order to keep the decay rate $\Gamma_2$ fixed. We fix the field strengths $F_2=\sqrt{2}F_1=2\cdot10^{-3}$ a.u. ($I_1=2I_2=1.4\cdot10^{11}$ W/cm$^2$) and the field polarizations as $\mathbf{e}_1=\hat{e}^L_+$, $\mathbf{e}_2=\hat{e}^L_+$, $\mathbf{e}_3=\hat{z}^L$, where $\mathbf{e}^L_{\pm}=(\hat{x}^L\mp\text{i}\hat{y}^L)/\sqrt{2}$. In Fig. \ref{Fig2} we assume that the axes of the molecular frame $(\hat{x}^M,\hat{y}^M,\hat{z}^M)$ are oriented along the corresponding laboratory ones $(\hat{x}^L,\hat{y}^L,\hat{z}^L)$. The transition dipoles are chosen as $\mathbf{d}_{1E}=\mathbf{e}^L_+$, $\mathbf{d}_{2E}=\mathbf{e}^L_+$ and $\mathbf{d}_{12}=\hat{z}^L$. The transition matrix elements are calculated as $M_{ij}=\mathbf{d}_{ij}^L\cdot\mathbf{F}^L$. Note that we assume that the each field couples only to the corresponding transition, as shown in Fig. \ref{Fig1}. This is a good approximation as long as the laser pulses bandwidth does not exceed the difference between the three transition frequencies.
\par For the above defined field strengths and dipole moments, we encircle the EP of the right enantiomer at the position $\Delta^{EP_R}=-1.77\cdot10^{-5}$ a.u. ($-4.83\cdot10^{-4}$ eV) and $F_3^{EP_R}=1.57\cdot10^{-6}$ a.u. ($I_3=8.66\cdot10^{4}$ W/cm$^2$). The path is parameterized as
\begin{eqnarray}
    \label{eq:path}
    \Delta(t)&=&x^{(0)}_0\pm\rho^{(0)}_x\sin(2\pi(t-t_0)/T_{loop})\\
    F_3(t)&=&y^{(0)}_0\pm\rho^{(0)}_y\cos(2\pi(t-t_0)/T_{loop})
\end{eqnarray}
where we choose $(x^{(0)}_0,y^{(0)}_0)=(\Delta^{EP_R},F^{EP_R}_3)$ and $\rho^{(0)}_x=|\Delta^{EP_R}|$, $\rho^{(0)}_y=|F_3^{EP_R}|$. The $\pm$ sign corresponds to the sense of encirclement. For all simulations reported in this work on topological population transfer we choose a loop time $T_{\mathrm{loop}}=3\cdot10^{5}$ a.u. (7.2 ps) to ensure that $T_{\mathrm{loop}}/\Delta>1$. The initial time $t_0$ is chosen such that $F_3(t_0)=F_3(t_0+T_{\mathrm{loop}})=0$. To quantify the efficiency of the asymmetric switch effect, we define the population inversion and $S$ parameters as defined in Ref. \cite{Feilhauer:2020aa}:
\begin{equation}
    A(t)=\frac{P_+(t)-P_-(t)}{P_+(t)+P_-(t)},
\end{equation}
\begin{equation}
    S=A(t=0)\cdot A(t=T_{\mathrm{loop}}).
\end{equation}
The $A(t)$ parameter quantifies the normalized population inversion between the adiabatic states of the molecule, while a $S$ parameter of 1 (-1) indicates that the population has been fully transferred at the end of the loop to the initially populated (depleted) state at the start of the dynamics.
We construct for each molecular enantiomer four total $A(t)$ and $S$ parameters, two per sense of encirclement and two per initial population in the $|\phi_+\rangle$ or $|\phi_-\rangle$ adiabatic state $\mathbf{A}(t)=\left[A^{\circlearrowleft}_+(t),A^{\circlearrowleft}_-(t),A^{\circlearrowright}_+(t),A^{\circlearrowright}_-(t)\right]$, $\mathbf{S}=\left[S^{\circlearrowleft}_+,S^{\circlearrowleft}_-,S^{\circlearrowright}_+,S^{\circlearrowright}_-\right]$. From these parameters, one can then construct the parameter $\alpha\in[-1,1]$ defined as
\begin{equation}
    \alpha=\left(S^{\circlearrowleft}_+S^{\circlearrowright}_+ + S^{\circlearrowleft}_-S^{\circlearrowright}_++S^{\circlearrowleft}_+S^{\circlearrowleft}_-+S^{\circlearrowright}_+ S^{\circlearrowright}_-\right)/4.
\end{equation}
If $\alpha<0$ the asymmetric switch effect is successful and independent of the initial conditions \cite{Feilhauer:2020aa}. We note that an asymmetric switch effect can still occur only for a given initial condition, in which case the $\alpha$ parameter might be non-negative. For the simulations in Figs. \ref{Fig3}, we deform the path defined in Eqs. \ref{eq:path} by replacing either $\rho^{(0)}_x$ with $\rho_x=\rho\cdot\rho^{(0)}_x$ ($\rho\in[0.25,2])$ or $x^{(0)}_0$ with $x_0=\delta x^{(0)}_0$ ($\delta\in[0,2])$. For all simulations we keep $T_{\mathrm{loop}}=3\cdot10^{5}~$ a.u..

To account for molecular orientations, we consider each orientation to be specified by three Euler angles with respect to the reference frame of the laboratory. We use a Lebedev grid of the 17th order for two of these angles, combined with a uniformly discretized grid with 360 points over the interval $[0,2\pi]$ for the remaining angle. For each orientation, we can calculate the corresponding $\alpha_{\Omega_i,\beta_j}$ parameter starting from the populations $|c_{\Omega_i,\beta_j}(t)|^2$ obtained from solving the TDSE in the molecular orientation specified by the $\Omega_i$ and $\beta_j$ from the discretized grid. The weight associated to each orientation is given by $w[\Omega_i]\Delta\beta$, where $w[\Omega_i]$ is the weight of the given point defined on the Lebedev grid, and $\Delta\beta=2\pi/360$. Given the populations $|c_{\Omega_i,\beta_j}(t)|^2$ for each orientation specified by the angles $\Omega_i$ and $\beta_j$, the orientation averaged population is $\langle |c(t)|^2\rangle=\sum_i\sum_j w_i\Delta\beta|c_{\Omega_i,\beta_j}(t)|^2/\sum_i\sum_jw_i\Delta\beta$. The orientation averaged $\langle\alpha\rangle$ parameters reported in the main text are calculated from the orientation averaged populations. For additional information on the details of the numerical simulations and further discussion on the effect of molecular orientation, we refer the reader to the SI \cite{SuppNotes}.

\section{Numerical details of the two-level chiral system}

We consider a circularly polarized vector potential in the laboratory frame $\mathbf{A}(t)=A_0\mathbf{e}_+\exp(\text{i}(kz-\omega t)$, where $\mathbf{e}_+=(\mathbf{e}_x-\text{i}\mathbf{e}_y)$, from which we derive the electric $\mathbf{E}(t)=-\partial_t\mathbf{A}(t)$ and magnetic $\mathbf{B}(t)=\nabla\times\mathbf{A}(t)$ fields. We assume that $A_0\omega=5\cdot10^{-4}$ a.u. and that the field is resonant with the transition so that $\Delta=0$. The chirality of the field is $\text{Im}\left[\mathbf{E}^*\cdot\mathbf{B}\right]\simeq1.8\cdot10^{-9}$ a.u.; the electric dipole in the molecular frame is $\mathbf{d}=\hat{e}^M_y$. If we assume the bound wavefunctions to be real the magnetic dipole is a purely imaginary pseudovector, which we take here to be $\mathbf{m}=\text{i}(\hat{e}^M_X+\hat{e}^M_y)/\sqrt{2}$. The chirality of the molecule is given by $\text{Im}\left[\mathbf{d}\cdot\mathbf{m}^*\right]=-0.707$. The TDSE is solved assuming the initial wavefunction to be in the lower energy state $|1\rangle$. For a given orientation we obtain the bound populations $P(t)=|c_1(t)|^2+|c_2(t)|^2$ for the two enantiomers and find the dichroism as $\text{CD}(t)=(P_R(t)-P_L(t))/(P_R(t)+P_L(t))$.

\section{Numerical details for the solution of fenchone molecules}

We perform \textit{ab-initio} quantum chemistry calculations using Gaussian16 \cite{g16} on the CREATE cluster \cite{kcl_create_2025} in fenchone at the B3LYP/aug-cc-pVDZ level using the optimized geometry at the MP2(fc)/cc-pVTZ level reported in Ref. \cite{Cireasa:2015aa}. At the sodium D-line (589 nm), we obtain an average molecular polarizability $\alpha_m=0.117488\cdot10^{3}$ a.u. (corresponding to $n=1.459$) and the specific optical rotation $[\alpha]=57.24\,\,\mathrm{deg}\,\mathrm{dm}^{-1}\mathrm{(g/mL)}^{-1}$. Both numbers are consistent with the experimental va\-lues reported in the literature \cite{Fenchone:SigmaAldrich}. The phase shift due to the optical rotation used in the simulations is given by $\alpha=[\alpha]\cdot\rho\cdot\Gamma_{ev}=2.104$ rad/m, where $\rho=0.948$ g/mL is the density of fenchone and $\Gamma_{\mathrm{evan}}$ is the fraction of the power in the evanescent modes. As electronic excitations for fenchone are in the UV range, the electronic circular dichroism at the sodium D-line is set to zero.

\section{Numerical details for the simulation of the chiral twisted fiber}

We consider the coupling of $\lambda=589$ nm light in a PMMA fiber with no cladding, with refractive index $n_{c}=1.4905$ \cite{Polyanskiy:2024aa}. For a solution of fenchone molecules with $n_s=1.459$ and a fiber radius of $r_{\mathrm{core}}=0.5\mu$m, the fiber supports a single mode with real propagation constant $\beta=0.984\cdot2\pi n_{c}/\lambda$. The electric field of the azimuthally symmetric $\mathrm{LP}_{01}$ mode is given by $E(\rho\le1)=A_0J_0(X\rho)$, $E(\rho>1)=A_0J_0(X)K_0(Y\rho)/K_0(Y)$, where $\rho=r/r_{\mathrm{core}}$ is the scaled radial coordinate and $X=r_{\mathrm{core}}\sqrt{n^2_{c}4\pi^2/\lambda^2-\beta^2}$ and $Y=r_{\mathrm{core}}\sqrt{\beta^2-n^2_{s}4\pi^2/\lambda^2}$. We find numerically the containment factor and corresponding power fraction in the evanescent modes as $\Gamma_{\mathrm{core}}\simeq0.833$, $\Gamma_{\mathrm{evan}}\simeq0.167$. The coupling term due to the torsion of the fiber is $k_t=\phi_t(1-R\cdot G\cdot n_{c})$ \cite{Sakai:1981aa}, where $R$ is the modulus of rigidity and $G$ is the photoelastic constant of PMMA. For the values reported in the literature it is safe to assume that $k_t\simeq\phi_t$. The input polarization in the fiber is RCP. The modes co-rotating with the fiber $\Psi(z)=\left(\Psi_-(z),\Psi_+(z)\right)$ are found by solving numerically the coupled modes equations. The total power is evaluated as $P(z)=|\Psi_+(z)|^2+|\Psi_-(z)|^2$. The ellipticity is $\xi(z)=\left(|\Psi_+(z)|^2-|\Psi_-(z)|^2\right)/\left(|\Psi_+(z)|^2+|\Psi_-(z)|^2\right)$, where $\xi=\pm1$ corresponds to RCP/LCP.

\bibliography{bibliography}% Produces the bibliography via BibTeX.

\end{document}